\documentclass[%
10pt,
 reprint,
superscriptaddress,
 amsmath,
 amssymb,
 aps,
 twocolumn
pra,
]{revtex4-2}
\usepackage{dcolumn}
\usepackage[utf8]{inputenc}
\usepackage{graphicx,bm,color}
\usepackage[dvipsnames]{xcolor}
\usepackage[margin=1in]{geometry}
\usepackage{amsthm,amsmath,amssymb,mathtools}
\usepackage{multirow}
\usepackage{todonotes}
\usepackage{tikz-cd} 
\usepackage{braket}
\usepackage[normalem]{ulem}
\definecolor{britishracinggreen}{rgb}{0.0, 0.5, 0.15}
\definecolor{burgundy}{rgb}{0.5, 0.0, 0.13}
\definecolor{egyptianblue}{rgb}{0.06, 0.2, 0.65}
\usepackage[colorlinks=true, citecolor=burgundy, linkcolor=britishracinggreen, urlcolor=egyptianblue]{hyperref}
\usepackage{tikz-cd}
\usepackage{enumitem}
\usepackage{ragged2e}
\usepackage[noend]{algpseudocode}
\usepackage{float}
\makeatletter
\let\newfloat\newfloat@ltx
\makeatother
\usepackage{algorithm}
\usepackage{multirow}
\usepackage{changes}
\definechangesauthor[name={Ilya Safro}, color=red]{is}
\definechangesauthor[name={Filip Maciejewski}, color=magenta]{fbm}
\makeatletter
\let\newfloat\newfloat@ltx
\makeatother
\usepackage{algorithm}
\usepackage{multirow}

\def\BibTeX{{\rm B\kern-.05em{\sc i\kern-.025em b}\kern-.08em
    T\kern-.1667em\lower.7ex\hbox{E}\kern-.125emX}}

\newcommand{\x}{\mathbf{x}}
\newcommand{\y}{\mathbf{y}}

\newcommand{\udel}{Department of Computer and Information Sciences, University of Delaware, Newark, DE, USA}

\newcommand{\purdue}{Davidson School of Chemical Engineering, Purdue University, West Lafayette, IN, USA}

\newcommand{\rigetti}{Rigetti Computing, Berkeley, CA}

\newcommand{\quail}{Quantum Artificial Intelligence Laboratory (QuAIL), NASA Ames Research Center, CA, USA}

\newcommand{\riacs}{USRA Research Institute for Advanced Computer Science (RIACS), CA, USA}
    
\begin{document}

\title{A Multilevel Approach For Solving Large-Scale QUBO Problems With Noisy Hybrid Quantum Approximate Optimization}

\author{Filip B. Maciejewski}
\email{fmaciejewski@usra.edu}
\affiliation{\quail}
\affiliation{\riacs}
\author{Bao Gia Bach}
\affiliation{\udel}
\author{Maxime Dupont}
\affiliation{\rigetti}
\author{P. Aaron Lott}
\affiliation{\quail}
\affiliation{\riacs}
\author{Bhuvanesh Sundar}
\affiliation{\rigetti}
\author{David E. {Bernal Neira}}
\affiliation{\riacs}
\affiliation{\purdue}
\author{Ilya Safro}
\affiliation{\udel}
\author{Davide Venturelli}
\affiliation{\quail}
\affiliation{\riacs}
\date{\today}

\begin{abstract}
Quantum approximate optimization is one of the promising candidates for useful quantum computation, particularly in the context of finding approximate solutions to Quadratic Unconstrained Binary Optimization (QUBO) problems.
However, the existing quantum processing units (QPUs) are relatively small, and canonical mappings of QUBO via the Ising model require one qubit per variable, rendering direct large-scale optimization infeasible.
In classical optimization, a general strategy for addressing many large-scale problems is via multilevel/multigrid methods, where the large target problem is iteratively coarsened, and the global solution is constructed from multiple small-scale optimization runs.
In this work, we experimentally test how existing QPUs perform when used as a sub-solver within such a multilevel strategy. 
To this aim, we combine and extend (via additional classical processing steps) the recently proposed Noise-Directed Adaptive Remapping (NDAR) and Quantum Relax $\&$ Round (QRR) algorithms. 
We first demonstrate the effectiveness of our heuristic extensions on Rigetti's superconducting transmon device Ankaa-2. 
We find approximate solutions to $10$ instances of fully connected $82$-qubit Sherrington-Kirkpatrick graphs with random integer-valued coefficients obtaining normalized approximation ratios (ARs) in the range $\sim 0.98-1.0$, and the same class with real-valued coefficients (ARs $\sim 0.94-1.0$).
Then, we implement the extended NDAR and QRR algorithms as subsolvers in the multilevel algorithm for $6$ large-scale graphs with at most $\sim 27,000$ variables. 
In practice, the QPU (with classical post-processing steps) is used to find approximate solutions to dozens of problems, at most $82$-qubit, which are iteratively used to construct the global solution.
We observe that quantum optimization results are competitive in terms of the quality of solutions when compared to classical heuristics used as subsolvers within the multilevel approach.\\
\end{abstract}
\maketitle

\section{Introduction}

Quadratic Unconstrained Binary Optimization (QUBO) problems are a powerful framework to formulate numerous industrially relevant computational challenges in various fields such as logistics~\cite{phillipson2024quantum}, finance~\cite{herman2023quantum}, aerospace applications~\cite{rieffel2024assessing}.
As such, much recent effort has been focused on empirically assessing whether quantum heuristics can provide any speedup over well-established classical heuristics \cite{osti2023shortdepthqaoa,pelofske2023scalingwholechipqaoahigherorder,sachdeva2024quantumoptimizationusing127qubit,dupont2023quantumenhanced, Dupont2024QRR,maciejewski2023design,abbas2023quantum,dupont2024lightcones,Ebadi2022MIS,Byun2022MIS,Kim2022MIS,King2023dwave5000spins,Nguyen2023rydberg}.
One of the approaches to optimizing QUBO problems with quantum processing units (QPUs) is quantum approximate optimization -- an umbrella term encompassing various heuristic techniques to find approximate solutions to (usually) combinatorial optimization problems.
This includes the original, well-known Quantum Approximate Optimization Algorithm (QAOA) \cite{farhi2014quantum}, but also a multitude of its variations, see, e.g., recent review \cite{abbas2023quantum}.

To solve a QUBO problem on a quantum device, one typically maps it to an instance of an Ising model that describes a two-body Hamiltonian constructed from Pauli $\sigma_{Z}$ operators.  
However, in the standard mapping, a single QUBO variable corresponds to a single qubit, restricting the usage of current quantum devices to solving only relatively small problems --  limited by the size of the quantum device, which typically falls into the range of dozens and hundreds of qubits.
Indeed, to the best of our knowledge, the biggest-scale experimental demonstrations of quantum approximate optimization for QUBO have been limited to at most few-hundred-variables sparse graphs in Refs.~\cite{osti2023shortdepthqaoa,pelofske2023scalingwholechipqaoahigherorder,sachdeva2024quantumoptimizationusing127qubit}, or few-dozen-variables dense graphs in Refs.\cite{dupont2023quantumenhanced, Dupont2024QRR,maciejewski2023design}, see also \cite{abbas2023quantum} for review.

In this work, we numerically and experimentally investigate the possibility of circumventing those scale limitations by employing a multilevel (MLVL) approach to solve QUBO problems using quantum approximate optimization.
We note that this is not the only known decomposition method that breaks down the global problem into smaller instances~\cite{dupont2024lightcones,tran2016hybrid}. 
Another related approach is to use dense encodings of the target variables~\cite{sciorilli2024towards, fuller2024approximate,tan2021qubit,sundar2024qubit}.

We summarize here our main novel contributions.
\begin{itemize}[leftmargin=0pt, labelwidth=1.5em, itemindent=1.5em, labelsep=0.5em]
\item We improve upon the recently proposed Noise-Directed Adaptive Remapping (NDAR) \cite{maciejewski2024ndar} and Quantum Relax \& Round (QRR) \cite{Dupont2024QRR} algorithms.
NDAR is a noise-tailored meta-algorithm involving an external loop, where each step requires running quantum optimization on a suitably gauge-transformed Hamiltonian.
QRR is a classical post-processing scheme that uses results of quantum approximate optimization to propose better solutions.
We combine and modify both algorithms by implementing a hardware-efficient ansatz (Time-Block QAOA from Ref.~\cite{maciejewski2023design}), by improving parameter optimization, and by augmenting with additional classical processing steps. Our improvements lead to at least $\sim$ 3x speedup and increased solution quality compared to prior work.

\item We improve and apply the methods introduced in recent works~\cite{angone2023hybrid,bach2024mlqaoa,ushijima2021multilevel}, where a novel multilevel solver for the Max-Cut (graph partitioning) problem was designed and numerically tested in small-scale QAOA simulations (IBM superconducting and DWave architectures). We find approximate solutions to $6$ large-scale QUBO problems (up to $\approx 27,600$ variables), including the first time the method was applied to problem instances with negative/positive weighted coefficients, using Rigetti's superconducting QPU Ankaa-2. We use the extended NDAR+QRR algorithm to solve the subsidiary QUBOs.
In practical terms, the QPU is used to find approximate solutions to multiple subsidiary instances of up to $82$-qubit problems, demonstrating a performance that is competitive with state-of-the-art classical heuristics within the multilevel setting. 
\end{itemize}

By improving and combining these two best-of-class approaches, our results are placed among the most complex experimental demonstrations of applied quantum optimization to date.

\section{Preliminaries}

\subsection{Problem formulations}\label{sec:preliminaries:problems}
In this work, we define an $n$-variable QUBO problem as \emph{maximization} over vector of binary variables $\x = \left(x_0, x_1, \dots, x_{n-1}\right)^{\mathrm{T}} \in \mathbb{B}^{n}$ (with $x_i \in \mathbb{B} = \left\{0,1\right\}$) specified by the real-valued, upper-triangular matrix $Q$ via $\max_{\x} \x^{\mathrm{T}}Q\x = \max_{\x} \sum_{i=0}^{n-1}\sum_{j=i}^{n-1} Q_{i,j} x_{i}x_{j}$.

An $n$-variable (or $n$-node) simple undirected graph Max-Cut problem is specified by the upper-triangular adjacency positive/negative edge matrix $W$  with zeros on the diagonal and involves \emph{maximization} of the cost function $\max_{\x} \sum_{i=0}^{n-1}\sum_{j=i}^{n-1} W_{i,j} \left(x_{i}+x_{j}-2x_{i}x_{j}\right)$.
An $n$-variable QUBO problem can be reformulated as a $\left(n+1\right)$-variable Max-Cut problem. 
Up to a multiplicative factor and a constant (irrelevant to optimization), this is done via mapping $W_{i,j} = - Q_{i,j}$ for $i\neq j$, and $W_{i,n} = Q_{i,i} + \sum_{j=i}^{n-1} Q_{i,j}$ for $i < n$ (edges between an additional node and rest of the graph), see, e.g., Refs.~\cite{Boros1991, Dunning2018mqlib}. 
If $\mathbf{y}$ is the corresponding Max-Cut solution, the solution to the original QUBO problem is recovered via $x_i = y_i \oplus_2 y_n$, where $\oplus_2$ denotes addition modulo 2.

An $n$-qubit Ising model is specified by Hamiltonian $H = \sum_{i=0}^{n-1} h_{i} \sigma_{Z}^{\left(i\right)}$ + $\sum_{i=0}^{n-1}\sum_{j=i}^{n-1} J_{i,j} \sigma_{Z}^{\left(i\right)} \sigma_{Z}^{\left(j\right)}$, where $\sigma_{Z}^{\left(k\right)}$ is a Pauli $\sigma_{Z}$ operator acting on $k$th qubit, $h_{k}$ and $J_{k,l}$ are real-valued coefficients called local fields and couplings, respectively. 
The corresponding Ising optimization problem is the \emph{minimization} of the Hamiltonian over classical states $\ket{\x} = \otimes_{i=0}^{n-1} \ket{x_i}$ (with $\ket{x_i}$ denoting standard computational basis states), as in $\min_{\ket{\x}} \bra{\x}H\ket{\x} = \min_{\x} \sum_{i=0}^{n-1} h_{i}\left(1-2x_i\right) + \sum_{i=0}^{n-1}\sum_{j=i}^{n-1} J_{i,j} \left(1-2x_i\right)\left(1-2x_j\right)$.
Note that since $\left(1-2x_{i}\right)\left(1-2x_{j}\right) = 1-2x_{i}-2x_{j}+4x_{i}x_{j} = 1-2\left(x_{i}+x_{j}-2x_{i}x_{j}\right)$, in this convention the Max-Cut optimization (maximization) corresponds to Ising optimization (minimization) by identifying $J_{i,j} = W_{i,j}$, and setting local fields to $0$.
Mapping of QUBO to Ising can be thus done, for example, by first mapping $n$-variable QUBO to $\left(n+1\right)$-variable Max-Cut and identifying it with corresponding $\left(n+1\right)$-qubit Ising Hamiltonian, and this is the approach we take in this paper. 

\subsection{Figures of merit}

To assess the quality of the solution, we compare the obtained cost $C_i$ to the size of the spectrum via renormalized approximation ratio (AR) (see, e.g., \cite{abbas2023quantum}) $\mathrm{AR}_i = \frac{C_i - C_{\min}}{C_{\max}-C_{\min}}$, where $C_{\min}$ and $C_{\max}$ are the smallest and the highest known values of the cost function.
Note that for maximization problems, this AR is equal to $1$ for the optimal solution and $0$ for the worst (minimal) solution.
Note that in the familiar case of unweighted Max-Cut, we have $C_{\min} = 0$, and the above is reduced to the standard approximation ratio.
For minimization, the AR is defined as $\mathrm{AR}_i = \frac{C_{\max} - C_i}{C_{\max}-C_{\min}}$.

\subsection{Multilevel solvers}\label{sec:preliminaries:ml_solvers}

A general strategy for addressing many large-scale computational problems on different hardware architectures, including various optimization problems on graphs, is using multilevel algorithms (also known as multiscale, multiresolution, and multigrid-inspired methods) \cite{brandt2003multigrid,safro2006graph,ron2011relaxation,leyffer2013fast,safro2011multiscale}. 
Specifically, in the quantum context, this multilevel framework has been explored for graph partitioning, clustering, and the Max-Cut problem, with the Quantum Approximate Optimization Algorithm (QAOA) serving as the main local processing component \cite{ushijima2021multilevel, angone2023hybrid, bach2024mlqaoa}. 
The motivation for combining the multilevel method with quantum optimization arises from the current state of quantum computers, which are constrained by a limited number of qubits.
The multilevel framework addresses this limitation by coarsening the original graph into a hierarchy of reduced-size coarser graphs, enabling the problem to be solved within the constraints imposed by quantum hardware. Each coarser graph approximates the previous finer one with respect to the optimization problem but requires fewer resources to solve it.

In essence, the multilevel approach begins by coarsening the original problem to create a series (also known as a hierarchy) of progressively simpler, related problems at coarser levels. At each coarse level $i$, the best-found solution serves as an initialization for the next finer solution at level $i-1$. This initialization is enhanced through what is commonly referred to as ``local processing'' (also known as a refinement), a cost-effective series of fast steps that involve only a few variables at a time but collectively revisit all variables of that level multiple times.
We refer the Reader to Ref.~\cite{brandt2003multigrid} for a pedagogical introduction to the subject.

Various coarsening-uncoarsening schedules exist in the multilevel algorithms to achieve a better optimization quality (e.g., W-cycle, and FMG ~\cite{brandt2003multigrid}), but in this work, we explore the most basic single V-cycle to minimize the effect of classical processing.
This setting involves sequentially generating a hierarchy of the next coarser graphs $\{G_l=(V_l, E_l, w_l)\}_{l=0}^L$, where $l$ is the index of the level, $G_0$ is the original large-scale graph, and $G_L$ is the coarsest graph. 
The coarsening process consists of (1) relaxation-based grouping pairs of nodes based on the recently introduced maximization version of the algebraic distance \cite{chen2011algebraic}  for graphs and (2) edge weight accumulation in each coarse level. 
Initially, every node is first placed in a random position on the surface of a $d$-dimensional sphere.
Following the initialization, several node-wise correction iterations are applied to maximize the total weighted distance between each node and its neighbors within the sphere, which maximizes the contribution of each node to the total energy of the system \cite{angone2023hybrid}. When convergence is reached, the embedding is finalized and the nodes are paired using a K-D tree. This matching constructs the coarser-level nodes that form the hierarchy.

After the hierarchy is created, the Max-Cut instance at the coarsest level is solved, and the solution is gradually interpolated level by level up to the finest level. 
At each level of the uncoarsening process, the $l$th level solution is initialized from level $l+1$ and further refined via sub-solvers (quantum or classical).
During this refinement stage, sub-problems are iteratively generated and solved.
Importantly, the maximal size of sub-problems (\textbf{MSS}) can be controlled to allow for implementation on limited-size QPU. 
If the solution of a generated sub-problem instance contributes to improving the final solution, a new sub-problem instance is produced and solved. 
The refinement stage ends when a specified maximal number of unsuccessful consecutive refinements (\textbf{MUR}) is reached.
Note that the MUR parameter allows the control of the total runtime of the algorithm to achieve quality/time trade-off, one of the main benefits of the multilevel algorithms.

\subsection{Quantum Approximate Optimization}\label{sec:preliminaries:qaoa}
\subsubsection{Quantum Approximation Optimization Algorithm}
While there are multiple approaches to quantum approximate optimization \cite{abbas2023quantum}, here we focus mainly on the canonical Quantum Approximate Optimization Algorithm (QAOA) introduced in 
 \cite{farhi2014quantum} and its extensions (described later in this section).
In this setting, the input state is $\ket{+}^{\otimes n}$ that is a tensor product of $+1$ eigenstates of Pauli $\sigma_{X}$ operator. 
The quantum circuit applied to the input state is constructed from parametrized mixer operator $\exp\left(-i\beta_{i}H_{M}\right)$ with Hamiltonian $H_{M} = \sum_{i=0}^{n-1}\sigma_{X}^{\left(i\right)}$ and phase separation operator $\exp\left(-i\gamma_{i}H\right)$, with $H$ being the Ising cost Hamiltonian (recall Section~\ref{sec:preliminaries:problems}).
In the above, $\gamma_i$ and $\beta_i$ are elements of $p$-dimensional real-valued parameter vectors $\boldsymbol\gamma$ and $\boldsymbol{\beta}$.
The quantum state is obtained via application of the unitaries of the form 
$\ket{\psi\left(\boldsymbol{\gamma}\boldsymbol{\beta}\right)} = \left(\prod_{i=1}^{p} \exp\left(-i\beta_{i}H_{M}\right) \exp\left(-i\gamma_{i}H\right)\right) \ket{+}^{\otimes n}$.
The aim of $p$-depth QAOA is to find $p$-dimensional $\mathbf\gamma$ and $\mathbf\beta$ that minimize the expected value of variational state evaluated on the cost Hamiltonian $\left<H\right>_{\gamma,\beta} = \bra{\psi\left(\boldsymbol{\gamma}\boldsymbol{\beta}\right)}H\ket{\psi\left(\boldsymbol{\gamma}\boldsymbol{\beta}\right)}$.
This can be achieved by using various parameter setting strategies, as used for classical black-box optimization.

\subsubsection{Noise-Directed Adaptive Remapping}
In this work, we implement an improved version of an extension of QAOA called Noise-Directed Adaptive Remapping (NDAR).
To the best of our knowledge, it is among the most performant quantum approximate optimization protocols to date, far outperforming standard QAOA in recent experimental demonstrations.
The method was introduced in Ref.~\cite{maciejewski2024ndar} for improved quantum optimization in the presence of certain types of hardware noise.
NDAR is a meta-algorithm applied on top of standard QAOA (or other quantum approximate optimization).
The main assumption of NDAR is that the noisy device has a special classical ``attractor'' state, towards which the dissipative processes push the overall system dynamics.
This attractor state is described by a bitstring $\ket{\x_{att}}$.
We note that this assumption was demonstrated experimentally to be a good approximation for superconducting devices, where the noise attractor is typically $\ket{\x_{att}} = \ket{0\dots 0}$ due to amplitude damping/dissipation \cite{maciejewski2024ndar}.
Each step of the NDAR loop involves performing quantum approximate optimization, identifying the best-cost bitstring, and re-mapping the cost Hamiltonian $H$ in a way that aligns $\ket{\x_{att}}$ with that best-found solution. 
This is done via bitflip transformations (also known as spin-reversal transforms \cite{boixo2013experimental,pelofske2019optimizing}) that exchange the definition of $\ket{0}$ and $\ket{1}$ from the point of view of the Hamiltonian.
Specifically, in each NDAR step, the gauge transformation specified by high-quality bitstring $\y$ is applied to the cost Hamiltonian $H \rightarrow H^{\y}$ (this is done in pre- and post-processing) in a way that $\bra{\y}H\ket{\y} = \bra{\x_{att}}H^{\y}\ket{\x_{att}}$.
This effectively causes the distribution of solutions in the current optimization to be centered around high-quality solutions from the previous step.
The iterative re-mapping of the Hamiltonian is done until the convergence criterion is met.
In Ref.~\cite{maciejewski2024ndar}, NDAR was experimentally demonstrated to highly outperform original QAOA in $82$-qubit experiments on fully-connected random graphs (so-called Sherrington-Kirkpatrick model).

\subsubsection{Time-Block QAOA Ansatz}
Another modification of the original QAOA we use is a Time-Block ansatz introduced in Ref.~\cite{maciejewski2023design} for hardware-efficient optimization. 
The Time-Block (TB) QAOA uses the SWAP network structure of typical QAOA circuits, which is necessary to implement phase separator operators of dense Hamiltonians on limited connectivity hardware.
A TB $k$-QAOA is a circuit ansatz constructed from a standard QAOA circuit by dividing it into batches of physical depth $k$ that are parametrized jointly and adding additional mixer operators between such new layers.
This effectively parametrizes a subset of interactions of the original Hamiltonian as a single-phase separator, like in standard QAOA.
Specifically, one full layer $\exp\left(-i\gamma_i H\right)$ is replaced by a ``partial layer'' $\exp\left(-i\gamma_i H_j\right)$, where $H_j$ contains a subset of interactions, and $\sum_{j=1}^{n/k} H_j = H$. 
The subsets of interactions in each $H_j$ are chosen by dividing the $n$ layers of entangling gates required to implement the full phase separator $\exp\left(-i \gamma_i H\right)$ into $\frac{n}{k}$ sets.

The Time-Block QAOA with small values of $k$ and $p$ offers shallow-depth variational circuits that can perform similarly to the original QAOA, with a gain of easier experimental implementation (see results in Ref.~\cite{maciejewski2023design}). 
Indeed, a depth $p$ Time-Block $k$-QAOA corresponds to a circuit of physical depth $\sim pk$, compared to $\sim nk$ for standard QAOA.

\subsubsection{Quantum Relax $\&$ Round}

Finally, the bit strings sampled from the quantum computers are used to estimate a two-point correlation matrix $\mathsf{Z}$ with entries $\mathsf{Z}_{ij}=(\delta_{ij}-1)\langle\sigma^{(i)}_z\sigma^{(j)}_z\rangle$, where $\delta_{ij}$ is the Kronecker delta. Then, an eigendecomposition of $\mathsf{Z}$ is performed, and its eigenvectors are sign-rounded, entrywise. Considering each of the sign-rounded eigenvectors as a candidate solution to the original problem, we compute their cost $C$ and keep the best one as the final solution.
This algorithm, known as the quantum relax-and-round (QRR) algorithm, was developed in  Ref.~\cite{Dupont2024QRR}. It was shown~\cite{Dupont2024QRR,dupont2024lightcones} that when sampling bit strings from the standard QAOA with $p$ layers, QRR converges asymptotically to the optimal solution with $p$. At $p=1$, the performance of QRR matches that of a classical relax-and-round algorithm performed on the adjacency matrix of a graph problem with entries $W_{ij}$, as defined in Sec.~\ref{sec:preliminaries:problems}. This was proven analytically in the large $n$ limit for several problem classes such as Sherrington-Kirkpatrick spin glasses, unit-weight random $3$-regular graphs, as well as circulant graphs, and supported numerically on other problems. We note that the resulting average performance at $p=1$ is much higher than that of the standard QAOA algorithm for the same quantum resources.

\section{Improved Noise-Directed Adaptive Remapping}\label{sec:extended_ndar}

In this section, we discuss our modifications to the original NDAR proposal and demonstrate their effectiveness in experiments on $82$ qubits on Rigetti's superconducting transmon device Ankaa-2.

\subsection{Modifications}

Recall that the original NDAR proposal consists of an external loop, where each iteration involves adaptively solving gauge-transformed cost function Hamiltonian via QAOA or another quantum approximate optimization method.
We modify the original NDAR algorithm by implementing the following changes:
\begin{enumerate}
[leftmargin=0pt, labelwidth=1.5em, itemindent=1.5em, labelsep=0.5em]
\item {\bf{(Pre-processing at iteration $0$)}} 
    NDAR at step $0$ implements quantum approximate optimization for the original Hamiltonian $H$. 
    Its representation is always provided with some implicit, ``default'' bitflip gauge choice. 
    Here, we generate $10^4$ random solutions, evaluate them on the cost Hamiltonian, and choose the best-cost solution $\y$ to specify the initial bitflip gauge.
    For the gauge-transformed Hamiltonian $H^{\y}$, the noise attractor state $\ket{0\dots 0}$ becomes a higher-quality solution, in a sense that $\bra{\y}H\ket{\y} = \bra{0 \dots 0}H^{\y} \ket{0\dots 0}$ (see \ref{sec:preliminaries:qaoa}).
\item {\bf{(Shallow-depth circuit ansatz)}} 
    Instead of using the original QAOA, here we implement a Time-Block $k$-QAOA from~\cite{maciejewski2023design}, see discussion in Section~\ref{sec:preliminaries:qaoa}.
    We choose $k=\in \left\{\frac{n}{10}, \frac{n}{5}\right\}$ to reduce physical depth approximately ten/five-fold compared to standard QAOA. 
    We set algorithmic depth (the number of parametrized layers) to $p=1$.
    \item {\bf{(Optimization over multiple gauges)}} We allow the optimizer to choose one of $4$ best-cost bitstrings from the previous step to specify allowed gauge transformations in the next step, as opposed to the single best bitstring in the original NDAR. 
    \item {\bf{(Quantum Relax $\&$ Round and extension)}} 
    At each iteration step, we take the results of quantum optimization and evaluate the QRR algorithm on them to find a better approximate solution (recall Section~\ref{sec:preliminaries:qaoa}).
    Moreover, we implement an alternative version of QRR that constructs weighted correlators of the form $J_{i,j}\left<\sigma_{z}^{\left(i\right)}\sigma_{z}^{\left(j\right)}\right>$, and otherwise works the same as QRR (see Section~\ref{sec:preliminaries:qaoa}.
    The algorithm, which we call w-QRR (weighted-QRR), aims to resemble the classical relax and round strategies.
    At each optimization, we implement both QRR and w-QRR on the QPU results and choose a better solution.
    \item {\bf{(Hamming Distance Quadratic Local Search)}} At the end of each step, we additionally explore the neighborhood of the best-cost bitstring by generating and evaluating all solutions within Hamming distance $2$. 
    We refer to this step as HDQLS.
\end{enumerate}

\begin{table*}[t!]
\begin{center}
\begin{tabular}{|c|c|c|c|c|c|}
\hline
\textbf{Graph name (U/W)} & \textbf{Id}   & \textbf{Nodes}         & \textbf{Edges (av. degree)}    & \textbf{Best cut (non-MLVL)} & \textbf{Ref.} \\ \hline
\textbf{soc-epinions (U)} & \textbf{s-e (U)}   & \multirow{2}{*}{26588} & \multirow{2}{*}{100120 (3.77)} & 70112  & \cite{sparsetamudataset2011} \\ \cline{1-2} \cline{5-6} 
\textbf{soc-epinions (W)}  & \textbf{s-e (W)}    &                        &                                & 11398   & here         \\ \hline
\textbf{Karloff 16-7-1 (U)} & \textbf{Krl (U)}   & \multirow{2}{*}{11440} & \multirow{2}{*}{3363360 (294)} & 2522520   & \cite{karloffgraphs1996}         \\ \cline{1-2} \cline{5-6} 
\textbf{Karloff 16-7-1 (W)} & \textbf{Krl (W)}        &                     &                                & 59685    & here         \\ \hline
\textbf{Rolling Stock Assignment 3 (W)}& \textbf{RLS3}  & 1750                   & 1530375 (874.5)       & 6829    & \cite{QUBOBenchmarkFraunhofer2024}         \\ \hline
\textbf{Queens 16 (W)} & \textbf{Q16}                   & 1035                   & 535095 (517)          & 14428  & \cite{QUBOBenchmarkFraunhofer2024}         \\ \hline
\end{tabular}
\caption{\label{table:graphs_global_summary} Large-scale graphs considered in this work, with reference solutions obtained via classical global heuristics \emph{not} employing our multilevel approach. 
"U" ("W") stands for "unweighted" ("weighted"), meaning equal (non-equal) weights at each edge. 
For the soc-epinions and the Karloff graph, the original graphs are unweighted and to obtain a weighted (general QUBO) version, we add random weights from $\left[-1,1\right]$ range. 
The two graphs from Ref.~\cite{QUBOBenchmarkFraunhofer2024} are already general QUBO -- here we map them to the Max-Cut problem (see Section~\ref{sec:preliminaries:problems}).
The "Best cut" column denotes the best-found cut using classical heuristics solvers from the MQLib library \cite{Dunning2018mqlib}.
Multiple heuristics are tested for each graph (see text description), and the shown results are the best cut found across at most $100$ random initialization of the solver (with an allowed running time of at most $15$ minutes each).
}
\end{center}
\end{table*}

\subsection{Experimental results on 82-qubit dense SK problems}

We now present the results of benchmarks of modified Noise-Directed Adaptive Remapping and compare it with the original version for the same $10$ instances of the Ising model as in Ref.~\cite{maciejewski2024ndar}.
Those Hamiltonians correspond to fully connected graphs, with integer-valued interactions taken randomly as $\pm 1$, so-called Sherrington-Kirkpatrick (SK) model \cite{sherrington1975solvable} on $82$ qubits.

We implement NDAR experiments with the adaptive optimizer known as Tree-Structured Parzen Estimators (TPE) \cite{hyperopt2013}, as implemented in package Optuna \cite{optuna2019}.
Each iteration step involves implementing Time-Block $8$-QAOA with TPE using $t=150$ cost function evaluations and gathering $s=1000$ samples at each evaluation, same as in Ref.~\cite{maciejewski2024ndar}.
Following Refs.~\cite{maciejewski2021modeling,maciejewski2024ndar}, we also allow the TPE to optimize over categorical variable controlling the gates' ordering of the ansatz.

\begin{figure}
    \centering
\includegraphics[width=0.45\textwidth]{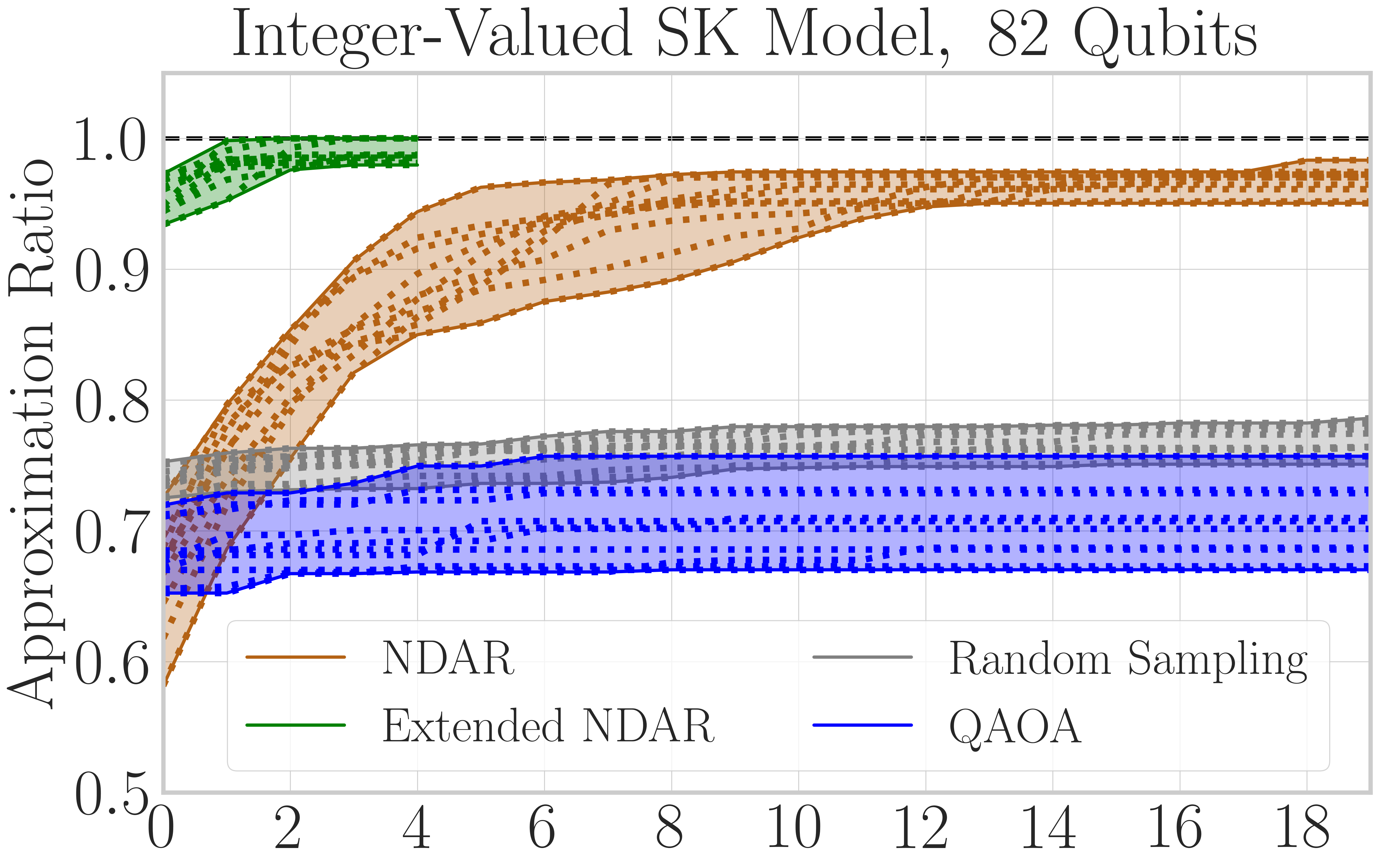}\\
\includegraphics[width=0.45\textwidth]{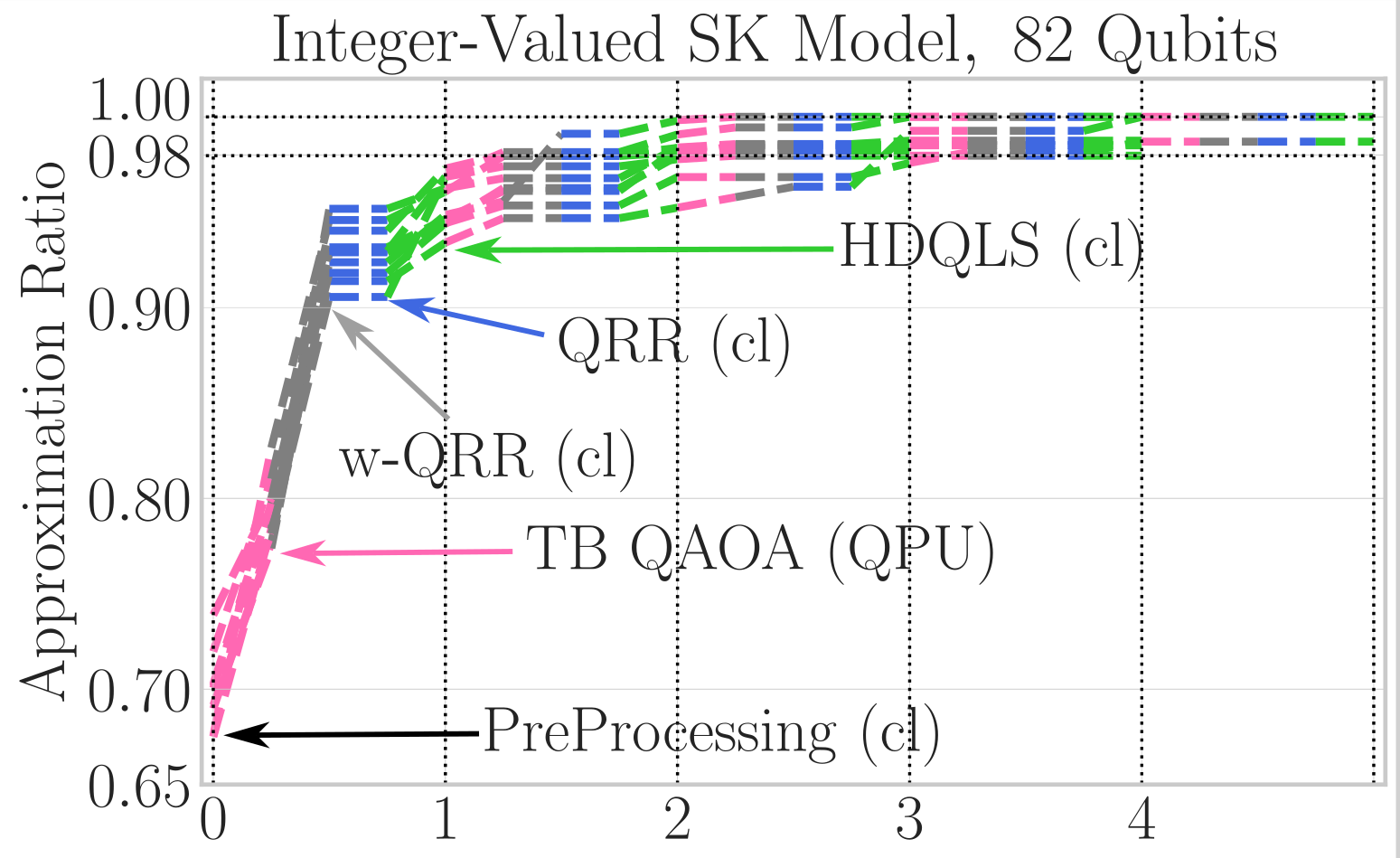}\\
\includegraphics[width=0.45\textwidth]{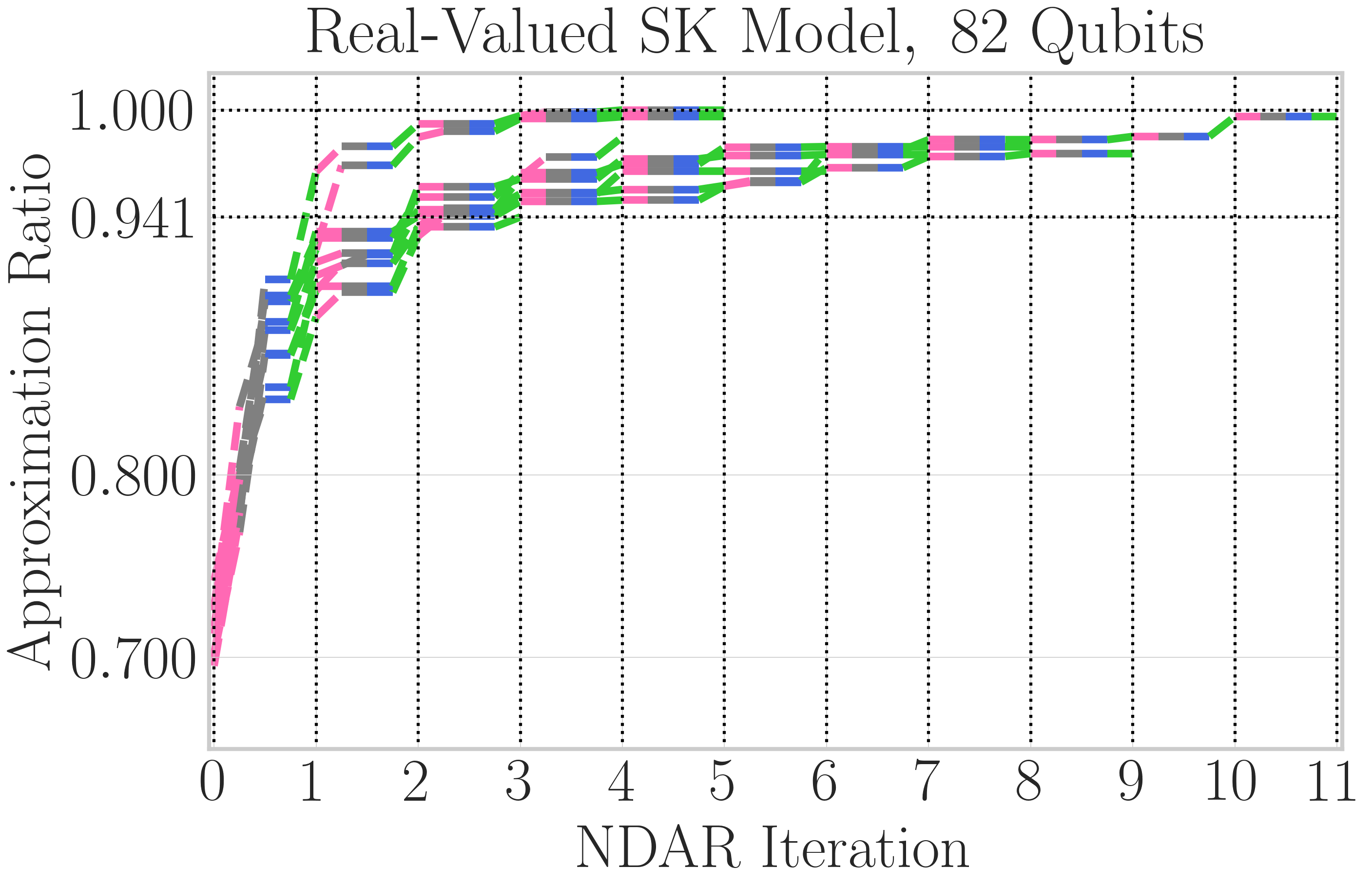}
\caption{
The top plot shows the experimental performance of extended NDAR (green), compared to original NDAR (orange), original QAOA (blue), and random bitstrings sampling (grey). 
The data points other than green are taken from Ref.~\cite{maciejewski2024ndar}.
Those experiments were performed on $10$ random instances of $82$-qubit integer-valued Sherrington-Kirkpatrick Hamiltonians.
The dotted lines within each colored region correspond to individual instances.
The middle plot shows a zoomed-in region of the top plot that corresponds to extended NDAR with more refined information. 
Here, the end of each colored region corresponds to AR, resulting from the application of the corresponding algorithm step (``cl'' stands for classical processing).
For example, the improvement gained directly from the QPU run (Time-Block QAOA) is indicated by the change in the AR value indicated by the pink line.
Different lines are different Hamiltonian instances.
Note that while the top plot shows the AR at the end of each optimization step, the middle plot shows refined information about ARs within each step. 
As such, AR at iteration $l$ on the top plot corresponds to $AR$ at the beginning of iteration $l+1$ (end of iteration $l$) in the middle plot.
The bottom plot shows the extended NDAR performance applied to real-valued Sherrington-Kirkpatrick models using the same convention.
Note that $y$-axis scales do not start at $0.0$ for improved clarity.
}\label{fig:exp:NDAR_vs_NDARplus}
\end{figure}

In the top plot of Fig.~\ref{fig:exp:NDAR_vs_NDARplus}, we present the joint result from Ref.~\cite{maciejewski2024ndar} and our implementation. 
We observe that the introduced modifications of the protocol lead to both better performance and much faster convergence of Noise-Directed Adaptive Remapping (at least $\sim$ 3x - although a rigorous parameter setting cost analysis should be conducted to determine the comparison~\cite{neira2024benchmarking}).
In the middle plot, we investigate in more detail how much each step of the extended NDAR (recall Section~\ref{sec:extended_ndar}) contributes to the final solution quality.
We implement QRR and w-QRR on the results from TB-QAOA, and HDQLS on the best bitstring obtained from this procedure.
We observe that the pre-processing allows one to start already from approximation ratios around $0.7$. Then, in the first iteration, the results obtained directly from QPU (via Time-Block QAOA) improve it to around $0.8$.
Weighted QRR provides further improvement of $0.1$, performing better than standard QRR (which does not improve upon w-QRR).
Finally, additional small improvements are gained by local search in the Hamming distance.
Further iterations of NDAR refine the solution quality until we obtain $\sim 0.98-1.0$ AR for all tested instances with at most $4$ NDAR iterations.

The bottom plot of Fig.~\ref{fig:exp:NDAR_vs_NDARplus} shows analogous investigation for $10$ random SK model instances with \emph{real-valued} interactions taken from range $\left[0,1\right]$.
We obtain approximation ratios in the range of $\sim 0.941-1.0$ after at most $10$ iterations. 
As such, the above results are placed among the most complex and most-performant applied quantum optimization experiments to date (see, e.g., Table~IV in Ref.~\cite{abbas2023quantum}).

\section{Multilevel approach to solving QUBO}

\subsection{Benchmark graphs}

In this work, we consider a set of $6$ benchmark graphs summarized in Table~\ref{table:graphs_global_summary}.
The biggest considered graph has $26588$ nodes with an average degree of $\sim 4$, while the densest graph is a $1750$-node problem with $\sim 875$ average degree.
We chose relatively small graphs compared to typical problems of industrial relevance so we could benchmark the multilevel method on the real QPU with limited resources.
The large-scale graphs are first solved using \emph{non-multilevel} (non-MLVL) heuristics chosen from the Mqlib library \cite{Dunning2018mqlib}, where the optimization is performed on the whole graph, as opposed to coarsened sub-problems generated by MLVL approach.
For each graph, we implemented around $25$ solvers (with a timeout of at most $15$ minutes) that use various heuristics including simulated annealing \cite{ALKHAMIS1999SA}, tabu-search  \cite{Beasley1998HeuristicAF,LU2010Tabu}, genetic algorithms \cite{Merz1999GLS}, and greedy local-search \cite{Merz2002GreedyLS}.
We note that the best or second-best solver for all graphs proved to be the Burer-Monteiro algorithm from Ref.~\cite{Burer2002}, which is a rank-2 relaxation of the famous Goemans-Williamson algorithm \cite{Goemans1995ImprovedAA}.

In Table~\ref{table:graphs_global_summary}, we show absolute values of the best cost function (cut) values found across all classical solvers.
Those values will be the reference when assessing the performance of the multilevel approach (via approximation ratio w.r.t. the non-MLVL optimizer) in the following sections.

\subsection{Classical heuristics sub-solver: MLVL algorithm}

We start by investigating the performance of the multilevel approach with classical subsolvers.
Recall from Section~\ref{sec:preliminaries:ml_solvers} that the MLVL algorithm from Ref.~\cite{angone2023hybrid} is specified by, among others, the following hyperameters -- Maximal Subproblem Size (MSS) and Maximal consecutive Unsuccessful Refinements (MUR).
MSS controls the biggest subproblems the solver is allowed to generate and thus can be set to the size of a given quantum device, allowing hybrid quantum-classical approximate optimization of large graphs using small-scale hardware.
MUR affects the run-time; it can thus be adjusted to the QPU availability.
In short, increasing MSS/MUR increases the space/time requirements of the multilevel approach.

We now numerically investigate how those parameters affect the optimization of our benchmark graphs when the sub-solver is \textit{classical} heuristics.
As sub-solver heuristics, we used $2$-$3$ solvers that performed the best in solving the original large graphs.
This included the Burer-Monteiro algorithm \cite{Burer2002} for all graphs, and other solvers \cite{ALKHAMIS1999SA, Beasley1998HeuristicAF, LU2010Tabu, Merz1999GLS, Merz2002GreedyLS} varied between graphs (see discussion in the previous Section); all sub-solvers were allowed to run for at most $10$ seconds per sub-problem.
The best-found results are presented in Table~\ref{table:graphs_ml_classical} (we note that the Burer-Monteiro algorithm proved to be usually the most performant also in the multilevel setting).
The mean approximation ratio is estimated empirically by running the solver for $10-100$ independent random initializations ($10$ for all cases except for the pair $\mathrm{MSS}=82$ and $\mathrm{MUR}=3$ with $100$ runs, where we increased statistics to have a better comparison with QPU implementation in the next section).
We observe a monotonic increase in solution quality for both hyperparameters.
We observe that for $5000$-variable subproblems, the multilevel solver starts to outperform global heuristics for the biggest considered graphs (soc-epinions), which corresponds to a five-fold size reduction compared to the original problem.
For other graphs, it becomes competitive around $500$-variable subproblems, except for the weighted Karloff graph where $2000$ variables are needed to obtain approximation ratios around $0.99$.
In the case of smaller QUBOs (RLS3 and Q16 from Ref.~\cite{QUBOBenchmarkFraunhofer2024}), the multilevel solver finds very good solutions already for $82$-node subproblems. 
However, in that case, it is worth noting that the algorithm's initial coarsening phase already finds high-quality solutions, and the subproblem refinements improve them only slightly. This indicates that the solutions of subproblems (either from classical or quantum sub-solvers) do not contribute much to the final solution.

\begin{figure}[!ht]
\centering
\includegraphics[width=0.45\textwidth]{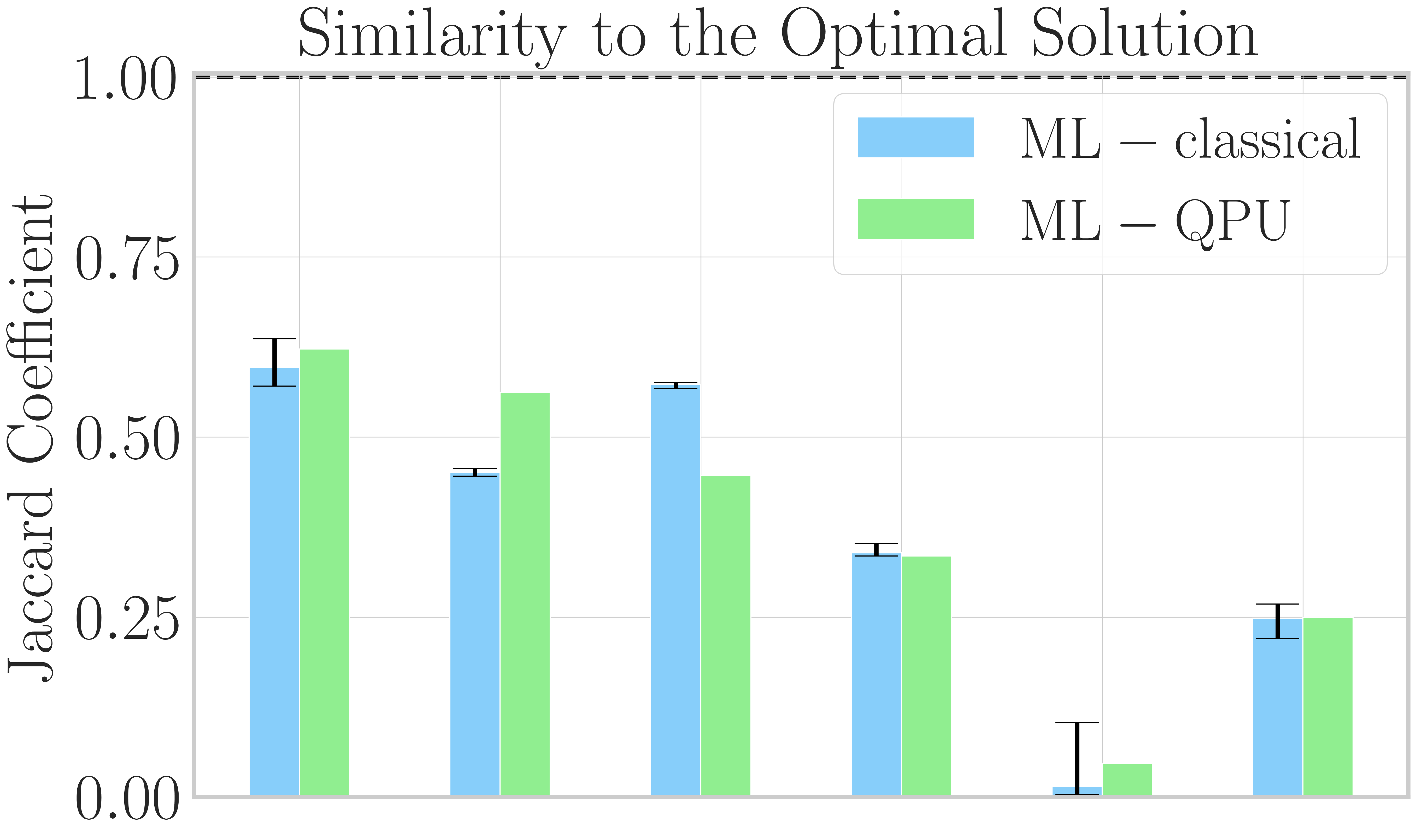}
\\
\includegraphics[width=0.45\textwidth]{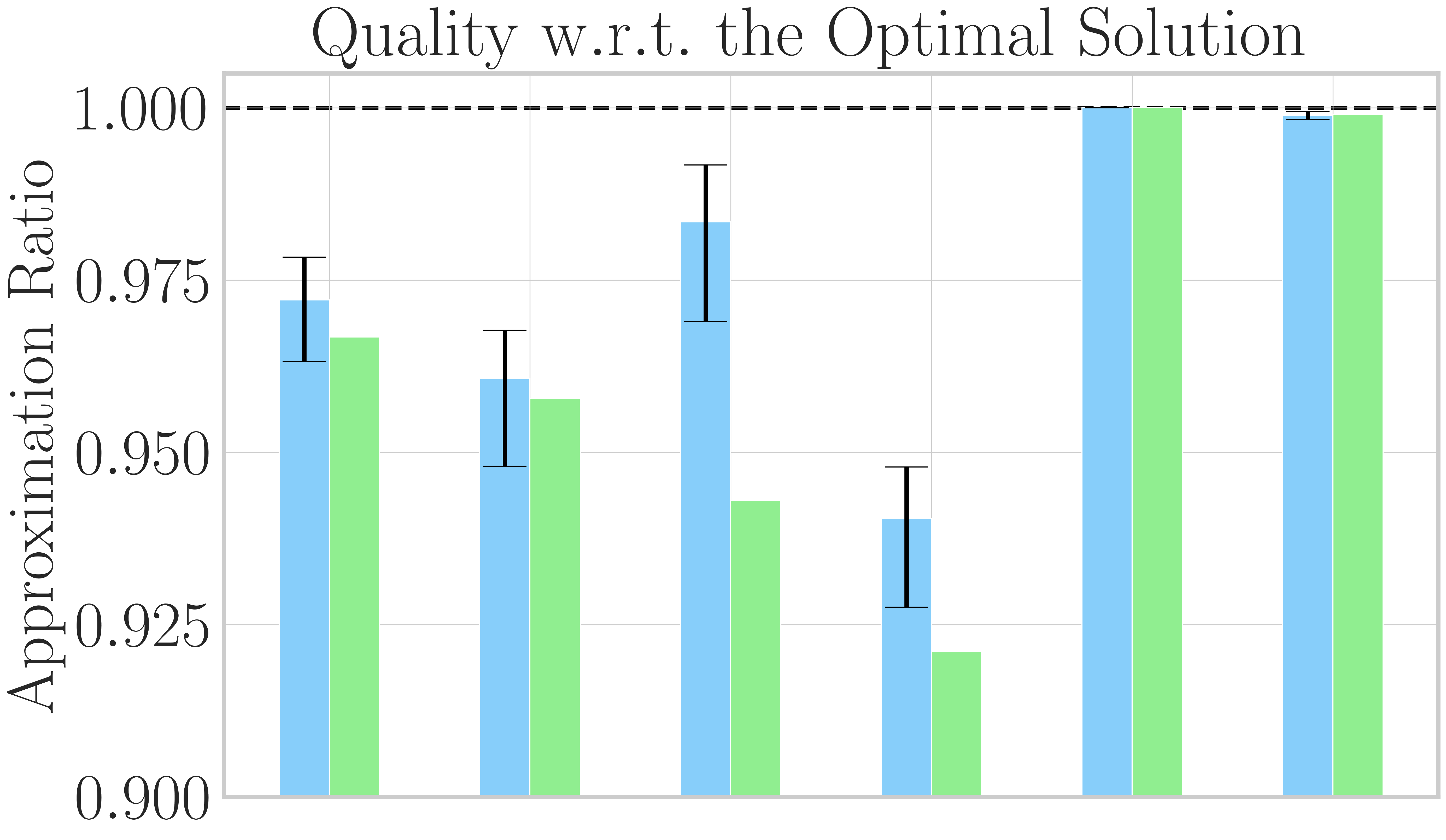}
\\
\includegraphics[width=0.45\textwidth]{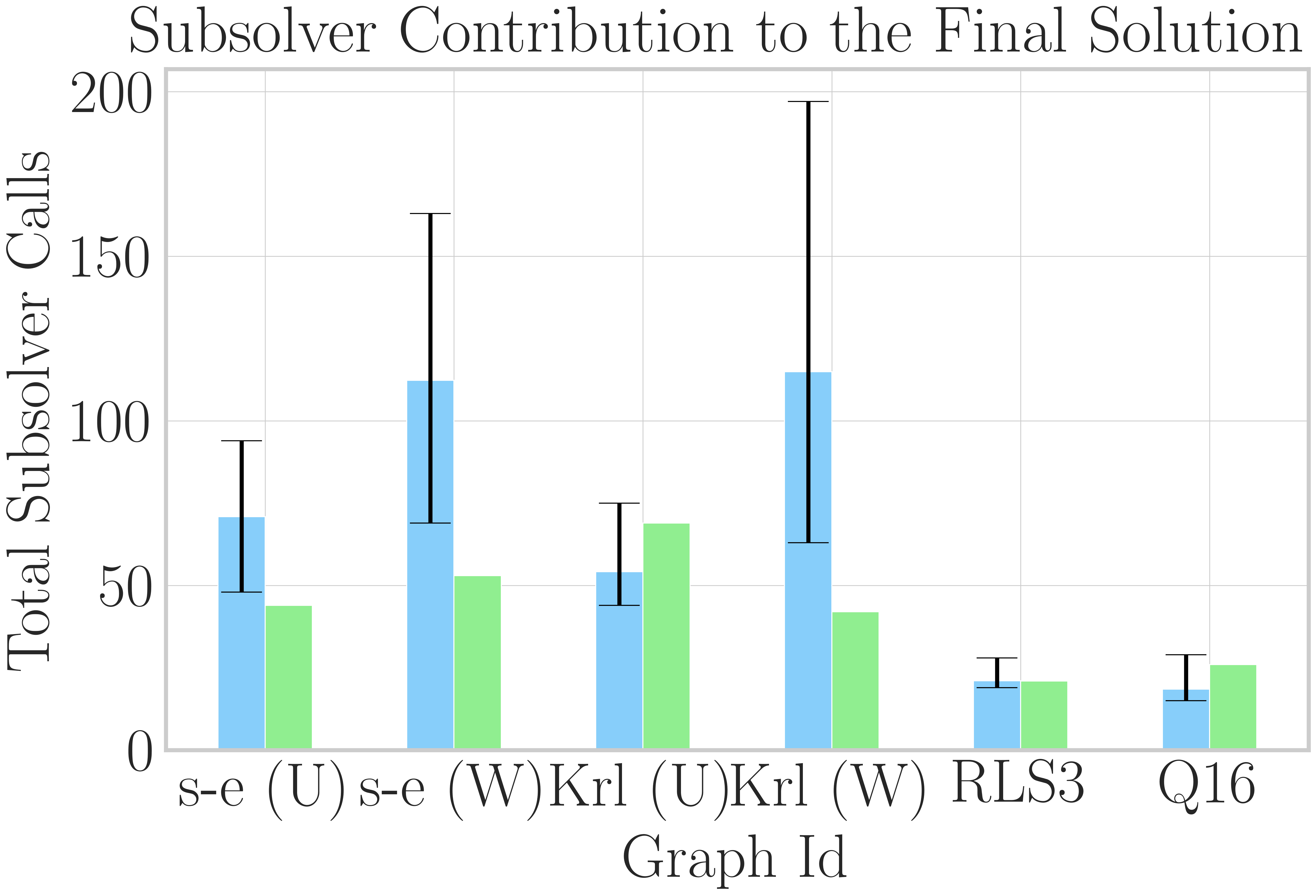}
\caption{
Results of Noise-Directed Adaptive Remapping implemented on Rigetti's QPU Ankaa-2 applied as sub-solver for the multilevel (ML) approach, compared to using classical subsolvers.
The hyperparameters are set to $\mathrm{MSS}=82$ and $\mathrm{MUR}=3$.
The top plot presents the Jaccard similarity coefficient between the edges participating in the cut for the optimal solution and the one from the multilevel approach.
The middle plot shows the approximation ratio (AR, note that $y$ scale starts at $0.9$ for improved visibility), while the bottom plot shows the total number of sub-solver calls in the multilevel approach.
For the classical approach, the error bars correspond to the range of obtained values (of best-found solutions) when running a multilevel algorithm with $100$ independent random initializations.
QPU results correspond to the best solution from a single experimental implementation.
We did not implement the MLVL method multiple times due to limited QPU availability.
}\label{fig:exp:ML_QPU_vs_classical}
\end{figure}

\subsection{82-qubit QPU NDAR sub-solver: MLVL algorithm}

For subproblems of size $82$ (the size of our experiments on Rigetti's QPU), the multilevel approach with classical sub-solver always finds worse solutions than non-MLVL heuristics, thus for those system sizes we do not expect to observe any advantage over those state-of-the-art heuristics.
Yet, it is worthwhile to benchmark how current QPUs perform compared to classical sub-solvers within the multilevel setting and possibly extrapolate the classical results from Table~\ref{table:graphs_ml_classical} to estimate what we can expect from future generations of QPUs. The objectives are to gain insights on the following questions in order to guide further R\&D: (i) How does the MLVL decomposition score against the best classical non-MLVL methods? (ii) are the solutions obtained by the quantum solvers similar to the ones obtained by the best classical method (multiscale or not) (iii) how does the performance when using the quantum method compare against a competitive non-quantum subsolver?
We implement the extended NDAR (recall Section~\ref{sec:extended_ndar}) as a multilevel sub-solver with at most $82$-qubit subproblems with  $\frac{n}{5}$ Time-Block ansatz. 
The results are presented in Fig.~\ref{fig:exp:ML_QPU_vs_classical}, together with corresponding results from Table~\ref{table:graphs_ml_classical} with classical subsolvers.

The top plot shows the Jaccard similarity coefficient (i.e., the size of the intersection divided by the size of the union) between the set of edges participating in a given cut (obtained via the multilevel approach) and the optimal cut (from non-MLVL solvers, see Table~\ref{table:graphs_global_summary}). 
We observe that the structural difference (measured by this coefficient) between solutions obtained from QPU and classical are minor, suggesting that QPU finds similar solutions to classical MLVL solvers. We note that their coefficients are significantly different from 1.0, which indicates that all solutions have structures distinct from those found in the best-found solution to the global problem. 

The middle plot shows the approximation ratio (compared to global solver solutions in Table.~\ref{table:graphs_global_summary}), while the bottom plot shows the total sub-solver calls, which gives an estimated runtime measured in the number of implemented local optimizations on $82$-variable subproblems.
We notice that for both unweighted and weighted soc-epinions, the QPU seems to converge faster than the classical sub-solver while offering solutions of similar quality.
For both Karloff graphs, QPU finds solutions with slightly worse ARs than the classical subsolvers while converging faster for the weighted variant.
For RLS3 and Q16 graphs, the QPU is similar in both quality and convergence speed.

We note that the ``number of subsolver calls'' metric used above treats the subsolver as a black box that, upon the call, returns an approximate solution to a given (at most) $82$-node subproblem.
As such, it does not differentiate between various levels of optimization complexity and provides only a rough comparison metric.
Here, in the classical case, the subsolver call involves running a classical heuristic for at most $10$ seconds on a standard laptop.
In the case of QPU, it involves running the whole NDAR+post-processing routine that typically ($\approx 80\%$ of datapoints) converged within $2-4$ iterations.
Each iteration involved implementing the Time-Block ansatz for $150$  different sets of variational parameters (trials), each trial involving gathering $10^3$ samples.
The QPU was thus used to gather between $3*10^5$-$6*10^5$ samples for each subproblem.
The typical $k=16$ Time-Block QAOA with $p=1$ circuit has a depth of around $65$ RX gates and $48$ ISWAP gates (see Refs.~\cite{maciejewski2023design,maciejewski2024ndar} for compilation techniques), which gives an estimate of around at most $\sim 10$s of pure QPU time per subproblem. 
We disregard here additional classical control overheads and classical optimizer updates, which can be high in practice but are expected to be significantly reduced for future pipelines.
The most expensive classical processing steps are QRR and its weighted version that for $n$-variable subproblem scale like $O\left(n^2\right)$ (see Ref.~\cite{Dupont2024QRR}), and quadratic Hamming distance search scaling as $O\left(n^2\right)$. %
We intend to explore more elaborate time-to-solution and energy-to-solution analyses in future work.

\begin{table}[h]
\begin{tabular}{|c|c|c|c|l|}
\hline
\textbf{Id}                        & \textbf{MSS}          & \textbf{MUR} & \textbf{MEAN} & \textbf{MAX} \\ \hline
\multirow{6}{*}{\textbf{s-e (U)}} &  \multirow{2}{*}{82}  & 3  & $0.972 \pm 0.008$  & 0.978 \\ \cline{3-5}
                                                &                       & 10  & $0.980 \pm 0.007$  & 0.983 \\ \cline{2-5}
                                                & \multirow{2}{*}{500}  & 3  & $0.982 \pm 0.009$  & 0.986 \\ \cline{3-5}
                                                &                       & 10  & $0.991 \pm 0.003$  & 0.993 \\ \cline{2-5}
                                                & \multirow{2}{*}{5000}  & 3  & $1.00 \pm 0.01$  & 1.001 \\ \cline{3-5}
                                                &                       & 10  & $1.001 \pm 0.001$  & 1.002 \\ \hline \hline
\multirow{6}{*}{\textbf{s-e (W)}} &   \multirow{2}{*}{82}  & 3  & $0.96 \pm 0.01$  & 0.968 \\ \cline{3-5}
                                                &                       & 10  & $0.969 \pm 0.007$  & 0.973 \\ \cline{2-5}
                                                & \multirow{2}{*}{500}  & 3  & $0.98 \pm 0.03$  & 0.990 \\ \cline{3-5}
                                                &                       & 10  & $0.992 \pm 0.008$  & 0.996 \\ \cline{2-5}
                                                & \multirow{2}{*}{5000}  & 3  & $0.99 \pm 0.02$  & 1.001 \\ \cline{3-5}
                                                &                       & 10  & $1.001 \pm 0.001$  & 1.002 \\ \hline \hline
\multirow{6}{*}{\textbf{Krl (U)}} &  \multirow{2}{*}{82}                    & 3  & $0.98 \pm 0.02$  & 0.992 \\ \cline{3-5}
                                                    &                       & 10  & $0.987 \pm 0.006$  & 0.990 \\ \cline{2-5}
                                                    & \multirow{2}{*}{500}  & 3  & $0.995 \pm 0.003$  & 0.997 \\ \cline{3-5}
                                                    &                       & 10  & $0.995 \pm 0.004$  & 0.997 \\ \cline{2-5}
                                                    & \multirow{2}{*}{2000}  & 3  & $0.999 \pm 0.002$  & 0.9994 \\ \cline{3-5}
                                                    &                       & 10  & $0.9996 \pm 0.0005$  & 0.9998 \\ \hline \hline      
\multirow{6}{*}{\textbf{Krl (W)}} & \multirow{2}{*}{82}                     & 3  & $0.94 \pm 0.01$  & 0.948 \\ \cline{3-5}
                                                    &                       & 10  & $0.950 \pm 0.007$  & 0.955 \\ \cline{2-5}
                                                    & \multirow{2}{*}{500}  & 3  & $0.96 \pm 0.01$  & 0.963 \\ \cline{3-5}
                                                    &                       & 10  & $0.976 \pm 0.003$  & 0.978 \\ \cline{2-5}
                                                    & \multirow{2}{*}{2000}  & 3  & $0.98 \pm 0.02$  & 0.984 \\ \cline{3-5}
                                                    &                       & 10  & $0.989 \pm 0.003$  & 0.991 \\ \hline \hline     
\multirow{4}{*}{\textbf{RLS3}} & \multirow{2}{*}{82}                     & 3  & $0.999990 \pm 0.000005$  & 0.999995 \\ \cline{3-5}
                                                &                       & 10  & $0.999991 \pm 0.000006$  & 0.999995 \\ \cline{2-5}
                                                & \multirow{2}{*}{500}  & 3  & $0.999996 \pm 0.000007$  & 0.999998 \\ \cline{3-5}
                                                &                       & 10  & $0.999998 \pm 0.000005$  & 0.9999999 \\ \hline \hline 
\multirow{4}{*}{\textbf{Q16}}               & \multirow{2}{*}{82}  & 3  & $0.9989 \pm 0.0009$  & 0.99946 \\ \cline{3-5}
                                            &                       & 10  & $0.9993 \pm 0.0007$  & 0.99970 \\ \cline{2-5}
                                            & \multirow{2}{*}{500}  & 3  & $0.99997 \pm 0.00007$  & 0.999997 \\ \cline{3-5}
                                            &                       & 10  & $0.99998 \pm 0.00004$  & 0.999997 \\   \hline  
\end{tabular}\caption{\label{table:graphs_ml_classical}
Results of classical heuristics applied as subsolvers for the multilevel approach. 
"MSS" stands for Max Subproblem Size, i.e., the maximal number of variables for each solved refinement subproblem.
"MUR" stands for "Max Unsuccesfull Consecutive Refinements", specifying how many non-improving refinements are performed before going to the next hierarchy level. See Section~\ref{sec:preliminaries:ml_solvers} for details.
For each set of MSS and MUR, each graph is solved using the multilevel approach with $10-100$ independent initialization for the refinements (the embedding is fixed).
The MEAN and MAX values refer to approximation ratios obtained in those independent runs.
Error bars indicate 3 standard deviations estimated empirically.
Approximation ratio (AR) is computed w.r.t. the results obtained with global solver, presented in Table~\ref{table:graphs_global_summary}.}
\end{table}

\section{Conclusions}
The ability to solve large-scale optimization problems on real quantum devices is crucial to demonstrating a practical quantum advantage in this area. 
Here, we present a complex hybrid quantum-classical algorithmic pipeline that is a candidate for achieving this goal. 
We introduced advanced modifications to our best-known approach for practical quantum optimization by hybridizing the Noise-Directed Adaptive Remapping (NDAR) \cite{maciejewski2024ndar} and Quantum Relax $\&$ Round (QRR) \cite{Dupont2024QRR} algorithms. 
We experimentally demonstrated that augmenting NDAR hardware-efficient Time-Block QAOA ansatz \cite{maciejewski2023design} runs with additional classical processing steps allows us to find high-quality solutions to fully-connected random graphs on $82$ qubits, improving upon previous experimental demonstrations.
Equipped with the new quantum heuristics, we investigated how they perform when used as a sub-solver in a multilevel approach to solving large-scale QUBO problems up to $\sim 27,000$ variables. 
\emph{Our results indicate that the real noisy QPU solver is competitive w.r.t. state-of-the-art heuristics used as sub-solvers in the multilevel approach.} 
Note that we restricted this initial work to analyze approximation ratios and solution similarity for a fixed meta-parameter setting strategy without estimating time-to-solution. Future work will further optimize NDAR and the MLVL scheme and will quantify additional metrics such as time-to-solution and energy-to-solution. We believe that our results constitute a promising step towards employing limited-scale, noisy quantum hardware for tackling industrially relevant large-scale problems.

\section*{Acknowledgments}
All authors acknowledge support by the Defense Advanced Research Projects Agency (DARPA) ONISQ program (for NASA/USRA: Agreement No. HR00112090058 and IAA8839, Annex 114). Authors from USRA also acknowledge support under NASA Academic Mission Services under contract No. NNA16BD14C. 
\bibliographystyle{unsrt}
\bibliography{bibliography}

\end{document}